\documentclass[12pt]{article}
\usepackage{amsmath}
\usepackage{latexsym}
\usepackage{amssymb}

\begin{document}

\title{A Pure Cotton Kink in a Funny Place\footnote{LMS Soliton meeting, Durham UK (Aug. 2004)} \footnote{PASCOS, Boston MA (Aug. 2004)
}}
\author{R. Jackiw\footnote{Electronic address: jackiw@lns.mit.edu}\\
{\small\itshape Center for Theoretical Physics}\\
{\small\itshape Massachusetts Institute of Technology}\\
{\small\itshape Cambridge, MA 02139-4307}}
\date{MIT-CTP-3485}
\maketitle


\thispagestyle{empty}

\begin{abstract}
When the 3-dimensional gravitational Chern-Simons term is reduced to two dimensions a dilaton-like gravity
theory emerges. Its solutions involve kinks, which therefore describe 3-dimensional, conformally flat spaces.

\end{abstract}
\maketitle

\pagestyle{myheadings}
\thispagestyle{myheadings}
Among the mathematical physics topics that have interested me over the years two are relevent
to the meeting at which I am now speaking: these are first, gravity theory and second, toplogical
entities both in mathematical settings like characteristic classes and in physical realizations
like kink profiles on a line. So for this event I shall present the results of an investigation
that unites these diverse elements.

Let me begin with lineal kinks. Consider the field equation
\begin{equation}
\Box \varphi - C \varphi + \varphi^3 = 0,
\label{varphi}
\end{equation}
where $C$ is a positive constant. When we look for a lineal kink solution, we take
$\varphi$ to depend on a single spatial variable. Then (\ref{varphi}) reduces to 
\begin{equation}
-\varphi'' - C \varphi + \varphi^3 = 0,
\label{-varphi}
\end{equation}
and has the well-known kink solution,
\begin{equation}
\varphi_k (x) = \sqrt{C} \tanh \sqrt{\frac{C}{2} x}, 
\label{frac}
\end{equation}
which interpolates between the ``vacuum" solutions $\varphi_0 = \pm \sqrt{C}$. \cite{rj1}
The kink has interesting roles in condensed matter physics where it triggers fermion
fractionization. \cite{rj2} Other kinks in other models give rise to completely solvable field theories, both
in classical and quantal frameworks. These  stories do not belong
here. But I shall return to the above kink later. 

Next let me consider the Chern-Simons characteristic class. In non-Abelian gauge
theory it is constructed from a matrix gauge connection $(A_\alpha)^\mu_{\ \nu}$ as \cite{rj3}
\begin{equation}
W(A) = \frac{1}{4 \pi^2} \, \int d^3 \, x \, \varepsilon^{\alpha \beta \gamma} \, t r
\bigg( \frac{1}{2} \, A_\alpha \, \partial_\beta \, A_\gamma \, + \, \frac{1}{3} \,
A_\alpha \, A_\beta \, A_\gamma\bigg).
\label{biggamma}
\end{equation}
This gauge theoretic entity finds physical application in the quantum Hall regime,
perhaps also in high $T$ superconductivity. When added with strength $m$ to the
usual Yang-Mills action, the Chern-Simons term gives rise to massive, yet gauge
invariant excitations in (2 + 1) - dimensional space-time. Also for consistency in the
quantized version of a non-Abelian theory $m$ must be an integer multiple of
$2\pi$. This is a precise field theoretic analog of Dirac's celebrated quantization of
magnetic monopole strength. Finally in an important mathematical application, the
Chern-Simons term gives  a functional integral formula for knot invariants.

The gauge theoretic  Chern-Simons term
(\ref{biggamma}) can be translated into a 3-dimensional geometric quantity by replacing the matrix gauge
connection
$(A_\alpha)^\mu_{\ \nu}$ with the Christoffel connection $\Gamma^\mu_{\alpha 
\nu}$. \cite{rj3}
\begin{equation}
W(\Gamma) = \frac{1}{4 \pi^2} \int d^3 x \  \varepsilon^{\alpha \beta \gamma} \
(\frac{1}{2}
\ \Gamma^\rho_{\alpha \sigma} \, \partial_\beta  {\Gamma^\sigma_{\gamma \rho}} + \frac{1}{3} 
{\Gamma}^\rho_{\alpha
\sigma} \, {\Gamma}^\sigma_{\beta \tau} \, \Gamma^\tau_{\gamma \rho}).
\label{2.1}
\end{equation}
But it is to be remembered that $\Gamma^\mu_{\alpha \nu}$ is constructed in the usual way from the
metric tensor, $g_{\mu \nu}$, which is taken as the fundamental, independent
variable. When $W(\Gamma)$ is varied with respect to $g_{\mu \nu}$ there emerges
the Cotton tensor, which has an important role in 3-dimensional geometry.
\begin{equation}
\delta W (\Gamma) = -\frac{1}{4\pi^2} \, \int d^3 x \delta g_{\mu \nu} \, \sqrt{g} \,
C^{\mu \nu}
\label{cottonten}
\end{equation}
\begin{equation}
C^{\mu \nu} \equiv \frac{1}{2\sqrt{g}} \bigg(\varepsilon^{\mu \alpha \beta} \,
D_\alpha \, R^\nu_\beta + \varepsilon^{\nu \alpha \beta} \, D_\alpha \,
R^\mu_\beta\bigg)
\label{tensorcott}
\end{equation}
[In (\ref{tensorcott}) one may freely replace the Ricci tensor $R^\mu_\nu$ by the
Einstein tensor $G^\mu_\nu \equiv R^\mu_\nu - \frac{1}{2} \, \delta^\mu_\nu \,
R^\alpha_{\ \alpha}$.] The Cotton tensor is like the covariant curl of the Ricci or
Einstein tensor. It is manifestly symmetric; it is covariantly conserved and traceless
because it is the variation of the diffeomorphism and conformally invariant $W(\Gamma)$. Furthermore,
the Cotton tensor replaces the Weyl tensor, which is absent in three dimensions, as a template
for conformal flatness: $C^{\mu \nu}$ vanishes if and only if the space is conformally
flat.
\begin{equation}
C^{\mu \nu} = 0 \Leftrightarrow \ \mbox{conformally flat space}
\label{leftright}
\end{equation}

Absence of the 3-dimensional Weyl tensor has the consequence that 3-dimensional geometries
satisfying Einstein's equation carry non-vanishing curvature only in regions where
there are sources. Therefore, there are no propagating excitations. However,
upon extending Einstein's gravity equation by adding $\frac{1}{m} \, C^{\mu \nu}$ to
the Einstein tensor [equivalently, adding $\frac{4\pi^2}{m} \, W(\Gamma)$ to the
Einstein-Hilbert action] the theory acquires a propagating mode with mass $m$, all
the while preserving diffeomorphism invariance! Here we have another perspective
on the absence of propagating modes in 3-dimensional Einstein theory: to regain the
Einstein equations from the modified equations, we must pass $m$ to infinity,
whereupon the super-massive propagating mode decouples.

This is a well known story, with which I do not concern myself now. Rather, I
consider the opposite limit of the extended theory, where only the Cotton tensor
survives, and the equation that I shall examine demands its vanishing, {\it i.e.} eq
(\ref{leftright}). But as indicated previously that equation is not sufficiently restrictive to be interesting:
any conformally flat space-time [coordinates $(t, x, y)$ ] is a solution. So I shall place a
further restriction: the solution that I seek should be independent of the {\it y} coordinate in a Kaluza-Klein
dimensional reduction from (2+1) to (1+1) dimensions of the gravitational Chern-Simons term $W(\Gamma)$
(\ref{2.1}) and of the Cotton tensor $C^{\mu \nu}$ (\ref{tensorcott}).
\cite{rj4}

To effect the dimensional reduction, We begin by making a Kaluza-Klein {\it Ansatz} for the 3-dimensional metric
tensor. It is taken in the form
\begin{equation}
3\mbox{-d} \, \mbox{metric tensor} = \varphi \left(
\begin{array}{cc}
g_{\alpha \beta} - a_\alpha a_\beta \quad &,\qquad -a_\alpha\\
-a_\beta \, &, \qquad -1 \quad
\end{array} \right),
\label{2.3}
\end{equation}
where the 2-dimensional metric  tensor $g_{\alpha \beta}$, vector $a_\alpha$, and scalar
$\varphi$ depend only on $t$ and $x$. [Henceforth, Greek letters from the beginning
of the alphabet $(\alpha, \beta, \gamma,...)$ index 2-dimensional ($t,
x$)-dependent  geometric entities, which are written  with lower case letters; in
three space-time dimensions geometric entities are capitalized (save the metric tensor)
and are indexed by middle Greek alphabet letters $(\mu, \nu, \rho...)$.]

It is easy to show that under infinitesimal diffeomorphisms, which leave the {\it y}-coordinate
unchanged, $g_{\alpha \beta}, a_\alpha   \ \mbox{and} \ \varphi$ transform as 2-dimensional coordinate tensor,
vector and scalar respectively, and moreover $a_\alpha$ undergoes a gauge transformation.

With the above {\it Ansatz} for the 3-dimensional metric, the Chern-Simons action becomes
\begin{equation}
CS = -\frac{1}{8 \pi^2} \int d^2 x \sqrt{-g} ( fr + f^3).
\label{2.4}
\end{equation}
Here $g= \mbox{det} g_{\alpha \beta}, \ \partial_\alpha a_\beta-\partial_\beta a_\alpha \equiv f_{\alpha \beta}
\equiv \sqrt{-g} \varepsilon_{\alpha \beta} f$, and $r$ is the 2-dimensional scalar curvature. The absence of a
$\varphi$ - dependence is a consequence of the conformal invariance of the gravitational Chern-Simons term, and
this also ensures that the Cotton tensor is traceless. Henceforth we set $\varphi$ to 1. 

The above expressions look like they
are describing 2-dimensional dilaton gravity, with $f$ taking the role of  a dilaton field.\cite{rj5} However, in fact $f$ is
not a fundamental field, rather it is the curl of the vector potential $a_\alpha$.  Alternative expressions for the
action (10) are
$\int d a (r + f^2)$, (where $d a$ is a 2-form; this exposes the topological character of our theory), and $ \int d^2 x
\Theta \ \varepsilon^{\alpha \beta} f_{\alpha \beta}, \ \Theta \equiv r + f^2$,  (which highlights an axion-like
interaction in 2-dimensional space-time).

Variation of $a_\alpha$ and $g_{\alpha \beta}$ produces the equations
\begin{eqnarray}
0&=&\varepsilon^{\alpha \beta} \partial_\beta\, (r + 3f^2), \label{2.5}\\
0&=&g_{\alpha \beta} \, (D^2 f-f^3 - \frac{1}{2} rf) - D_\alpha D_\beta f.
\label{2.6}
\end{eqnarray}
The first is solved by
\begin{equation}
r +3 f^2 = \mbox{constant} \equiv C.
\label{2.7}
\end{equation}
Eliminating $r$  in the second equation, and decomposing it into the trace and trace-free parts leaves
\begin{eqnarray}
0&=& D^2 f - Cf + f^3, \label{2.8}\\
0&=& D_\alpha D_\beta f -\frac{1}{2} g_{\alpha \beta} D^2 f.
\label{2.9}
\end{eqnarray}
Note that the equations are invariant against changing the sign of $f$ (the action then also changes sign). 

A
homogenous solution that respects the $f \leftrightarrow -f$ symmetry is 
\begin{equation}
f = 0, \ r=C.
\label{2.10}
\end{equation}
However, there also is a ``symmetry breaking" solution. 
\begin{equation}
 f= \pm \sqrt{C}, \ r = - 2C \ \ C>0
\label{2.11}
\end{equation} 

Forms for $g_{\alpha \beta}$ and $a_\alpha$ that lead to the above results are
\begin{eqnarray}
\mbox{(a)}& f&=\ 0, \ r \ = \ C > 0: g_{\alpha \beta} = \frac{2}{Ct^2}
\left[
\begin{array}{cc} 
1&0\\
0&-1
\end{array} \right],\quad a_\alpha = (0,0), \label{2.12}\\ [8pt]
\mbox{(b)}& f&=\ 0, \ r \ = \ C < 0: g_{\alpha \beta} = \frac{2}{|C| x^2}
\left[
\begin{array}{cc} 
1&0\\
0&-1
\end{array} \right],\quad a_\alpha = (0,0), \label{2.13}\\[8pt]
\mbox{(c)}& f&=\ \pm \sqrt{C}, \ r \ = \ -2C < 0: g_{\alpha \beta} = \frac{1}{C x^2}
\left[
\begin{array}{cc} 
1&0\\
0&-1
\end{array} \right], \quad a_\alpha = \left(\frac{\mp 1}{\sqrt{C} x}, 0 \right). \qquad
\label{2.14} 
\end{eqnarray}

In the first case, the 2-dimensional space-time is deSitter; in the last two, it is anti-deSitter.

The 3-dimensional scalar curvature $R$, with metric tensor as in our {\it Ansatz} (\ref{2.3}) at $\varphi=1$, is
related to the 2-dimensional curvature $r$ by
\begin{equation}
R=r+\frac{1}{2} f^2.
\label{2.21}
\end{equation}
Hence for the three cases, the 3-dimensional curvature and line element read
\begin{eqnarray}
\mbox{(a)}& R&=\ C>0,  \  \ (ds)^2 = \frac{2}{C}
\left[ \left(\frac{dt}{t}\right)^2 - \left(\frac{dx}{t}\right)^2 \right] -(dy)^2, \label{2.16}\\ [8pt]  
\mbox{(b)}& R&=\ C<0,  \  \ (ds)^2 = \frac{2}{|C|}
\left[ \left(\frac{dt}{x}\right)^2 - \left(\frac{dx}{x}\right)^2 \right] -(dy)^2, \label{2.17}\\[8pt]
\mbox{(c)}& R&=\ -\frac{3}{2}C<0,  \ \ (ds)^2 = \pm \frac{2}{\sqrt{C} x}
dtdy - \left(\frac{dx}{\sqrt{C}x} \right)^2- (dy)^2.
\label{2.18}  
\end{eqnarray}

Although all three solutions carry constant 3-dimensional curvature, the ``symmetry breaking" solution, (c) above, possesses greater geometrical symmetry in 3-dimensions: it supports 6 Killing vectors that span
$S O \mbox{(2.1)} \times S O \mbox{(2.1)} = S O \mbox{(2.2)}$, the isometry of 3-dimensional anti-deSitter space. Moreover one verifies that,  as expected,
$R^\mu_\nu = \frac{1}{3} \delta^\mu_\nu \, R= -\frac{1}{2} \, \delta^\mu_\nu \, C$ -- the 3-dimensional space-time
is maximally symmetric. The ``symmetry preserving" solutions, (a) and (b) above, admit only 4 Killing vectors that
span $S O
\mbox{(2.1)}
\times S O \mbox{(2)}$. 

Since the Cotton tensor vanishes, we expect that the above
space-times, are (locally) conformally flat. This can be seen explicitly for the ``symmetry preserving" solutions. In
(\ref{2.16})
$\mbox{set} \ T  = t \cosh \sqrt{\frac{C}{2}} y, \ Y = t \sinh \sqrt{\frac{C}{2}}y,\ \mbox{and} \ X=x \ \mbox{to find}$
\begin{eqnarray}
 \ (ds)^2 = \frac{2}{C (T^2 - Y^2)} \bigg((d T)^2 - (dX)^2 - (dY)^2 \bigg).
\label{2.19}
\end{eqnarray}
while in (\ref{2.17}) the coordinate transformation $X = x \cos \sqrt{\frac{|C|}{2}} y, \ Y= x \sin \sqrt{\frac{|C|}{2}} y, \
T = t$ gives the line element
\begin{equation}
(ds)^2 = \frac{2}{|C| (X^2 +Y^2)} \ \bigg((d T)^2 - (dX)^2 - (dY)^2 \bigg).
\label{2.20}
\end{equation}
The relevant transformation, which takes the ``symmetry breaking" solution into a conformally flat coordinate has been found by two graduate students. \cite{rjnew6}\\ It is given by
\begin{eqnarray}
T + Y = t + x \tanh \frac{\sqrt{C}}{2} y,\ T - Y = - \frac{1}{\sqrt{C}} \, \tanh \frac{\sqrt{C}}{2} \, y, \nonumber \\
X = \sqrt{\frac{x}{\sqrt{C}}} \, \frac{1}{\cosh \frac{\sqrt{C}}{2} y} \qquad \qquad \qquad \qquad \qquad \qquad \quad \ \ \ \,
\label{neweq}
\end{eqnarray}
Then the line element (24) becomes
\begin{equation}
(ds)^2 = \frac{4}{C X^2} \, \bigg((dT)^2 - (dX)^2 - (dY)^2\bigg)
\label{neweq2}
\end{equation}
and puts into evidence the anti-deSitter geometry.

The equations (\ref{2.8}), (\ref{2.9}) also posses a kink solution, which interpolates between the ``symmetry
breaking" solutions (\ref{2.11}). One can verify that
\begin{equation}
f (x) = \sqrt{C} \ \tanh \frac{\sqrt{C}}{2} \ x,
\label{2.21}
\end{equation}
with
\begin{equation}
\qquad \qquad   g_{\alpha \beta} = \left(
\begin{array}{cc}
1/ \cosh^4 \frac{\sqrt{C}}{2} x & 0\\
0& -1
\end{array}
\right),
\label{2.22} 
\end{equation}
satisfies the relevant equations. That the solution depends only on one variable (only $x$, not both $t, x$) is a general property (provided coordinates are selected properly). Thus in (\ref{2.8}), (\ref{2.9}) one is dealing with a system 
of second-order ordinary (not partial) differential equations, whose solution involves two integration constants. One
integration constant is the trivial origin of the $x$ coordinate [taken to be $x=0$ in(\ref{2.21}), (\ref{2.22})].  The other involves
choosing an integration constant in a first integral, so that one achieves a kink: a profile that interpolates between
$\pm \sqrt{C}$ as $x \to \pm \infty$. (Other choices for this second constant lead to the same local geometry, but to
different global properties. This has been thoroughly explained by Grumiller and Kummer. \cite{rj6}) 

The 2-dimensional curvature corresponding to (\ref{neweq2}) is 
\begin{equation}
r = -2C + \frac{3C}{\cosh^2 (\frac{\sqrt{C}}{2} x)}.
\label{2.23}
\end{equation}
Also the 3-dimensional line element, for (\ref{neweq}), (\ref{neweq2}) reads
\begin{equation}
(ds)^2 = -(dx)^2 -\frac{2}{\cosh^2 \frac{\sqrt{C}}{2} \ x}\, dtdy -(dy)^2,
\label{2.24}
\end{equation}
and the 3-dimensional scalar curvature is according to (21), (\ref{neweq}) and (\ref{neweq2})
\begin{equation}
R = -\frac{3C}{2} + \frac{5C}{2 \cosh^2 \frac{\sqrt{C}}{2} \ x}.
\label{2.25}
\end{equation}
The 3-dimensional scalar curvature (\ref{2.25}) clearly tends at large $x$ to its ``symmetry breaking" value (\ref{2.18}), but the relation of the line element (\ref{2.24}) to the one on (\ref{2.18})  is not evident. To expose it, we change coordinates in (\ref{2.24}) by
\begin{equation}
\bar{t} = t + y/2 \ , \ \bar{x} = \frac{1}{\sqrt{C}} \sinh^2 \frac{\sqrt{C}}{2} \, x, \ \bar{y} = y
\label{neweq3}
\end{equation}
Then the ``kink" line element (\ref{2.24}) becomes conformal to the ``symmetry breaking" line element (\ref{2.18}).
\begin{equation}
(ds)^2_{kink} \ = \frac{\bar{x}}{\bar{x} + 1/\sqrt{C}}\ \Bigg(\pm\frac{2}{\sqrt{C} \, x}\, d \bar{t} d \bar{y}  - \bigg(\frac{d\bar{x}}{\sqrt{C}\, \bar{x}}\bigg)^2  -(d \bar{y})^2\Bigg)
\label{neweq4}
\end{equation}
It is now clear that at large $\bar{x}$ the ``kink" and the``symmetry breaking" 
line elements coincide, while at finite $\bar{x}$, the kink provides a deformation of anti-deSitter space.
Expression (\ref{neweq4}) for the ``kink" line element, together with the previously given 
transformation (\ref{neweq}) [replace $(t, x, y)$ by $(\bar{t}, \bar{x}, \bar{y})$] determines a
conformally flat ``kink" line element.
\begin{equation}
(ds)^2_{kink} = \frac{4}{1-C\, (T-Y)^2 + C X^2} \bigg((dT)^2 - (dX)^2 - (dY)^2\bigg)
\label{eq36}
\end{equation}
Once again we see that (\ref{eq36}) tends to (\ref{neweq2}) at large $X$.

This then is the kink in a ``funny place" -- in conformally flat (2+1)-dimensional space-time. A question remains: can
one understand a {\it priori} that such a kink should exist in that geometry?
This question may be posed in a more general setting.

Observe that the flat space kink in Eqs. (\ref{varphi})-(\ref{frac}), possesses the same profile as (\ref{2.21}), except
for a change in scale. In fact this is a general feature. The following can be proven. If the non-linear equation in flat
space-time,
\begin{equation}
\Box\, \varphi + V'\, (\varphi) =0,
\label{boxvar}
\end{equation}
possesses a kink solution $\varphi_k \, (x) = k \, (x)$, then the curved (1+1)-dimensional space-time equations
\begin{equation}
D^2 \, f + V' \, (f) = 0,
\label{vone}
\end{equation}
\begin{equation}
D_\alpha \, D_\beta \, f - \frac{1}{2} \, g_{\alpha \beta} \, D^2 \, f = 0,
\label{threefour}
\end{equation}
are solved by
\begin{equation}
f (x) = k ( x/ \sqrt{2}),
\label{2.31}
\end{equation}
with 2-dimensional line element
\begin{equation}
(ds)^2 = \mbox{V} (f) (dt)^2 - (dx)^2,
\label{2.32}
\end{equation}
leading to a 2-dimensional curvature
\begin{equation}
r = - \mbox{V}^{\prime \prime} (F).
\label{2.33}
\end{equation}

In another generalization one can consider the supersymmetric extension of the Chern-Simons section 
and Cotton tensor. [8] In a dimensional reduction, similar to the one discussed previously, one is 
led to a supersymmetric generalization of  (10). Then the profiles that we have found, together with 
vanishing values for the fermionic fields, continue to be solutions of the supersymmetric equations. 
One may also determine the supersymmetry properties of our solutions. One finds that the 
``symmetry preserving" solutions (\ref{2.10}), (\ref{2.12}) and (\ref{2.13}) are not preserved by any of the supersymmetric transformations. 
The ``symmetry breaking" solution (\ref{2.11}), (\ref{2.14}), as well as the kink solution (\ref{2.21}), (\ref{2.22}) preserve half the of
supersymmetries. \cite{rj9} Furthermore, it is interesting to notice that  our model has also emerged in a recent study of BPS
solutions to $N=2$, $D=4$ gauged supergravity. \cite{rjr}

\end{document}